\documentclass[aip,pop,preprint,numerical,onecolumn,superscriptaddress]{revtex4}
\usepackage{amssymb, amsmath,framed}

\usepackage{bm}
\usepackage[colorlinks=true,linkcolor=blue]{hyperref}
\expandafter\ifx\csname package@font\endcsname\relax\else
 \expandafter\expandafter
 \expandafter\usepackage
 \expandafter\expandafter
 \expandafter{\csname package@font\endcsname}
\fi
\def\bq{\begin{equation}}
\def\eq{\end{equation}}
\def\bqy{\begin{eqnarray}}
\def\eqy{\end{eqnarray}}

\def\de{\delta}

\def\p{\partial}

\def\rh{\rho}

\def\p{\partial}

\def\cala{\mathcal{A}}
\def\calb{\mathcal{B}}
\def\calc{\mathcal{C}}

\def\calj{\mathcal{J}}

\def\calp{\mathcal{P}}

\begin{document}

\title{Remarkable connections between extended magnetohydrodynamics models}

\author{M.~Lingam}
\email{manasvi@physics.utexas.edu}
\author{P.~J.~Morrison}
\email{morrison@physics.utexas.edu}
\author{G.~Miloshevich}
\email{gmilosh@physics.utexas.edu}
\affiliation{Department of Physics and Institute for Fusion Studies, The University of Texas at Austin, Austin, TX 78712, USA}

\date{}

\begin{abstract}
Through the use of suitable variable transformations, the commonality of all extended magnetohydrodynamics (MHD) models is established. Remarkable correspondences between the Poisson brackets of inertialess Hall MHD and inertial MHD (which has electron inertia, but not the Hall drift) and extended MHD (which has both effects), are  established.  The helicities (two in all) for each of these models are obtained through these correspondences.  The commonality of all the extended MHD models is traced to the  existence of two Lie-dragged 2-forms, which are closely associated with the canonical momenta of the two underlying species.  The Lie-dragging of these 2-forms by suitable velocities also leads to the correct equations of motion. The Hall MHD Poisson bracket is analyzed in detail, and the Jacobi identity is verified through a detailed proof and this proof ensures the Jacobi identity for the Poisson brackets of all the models. 
\end{abstract}

\maketitle



\section{Introduction}
 \label{Intro}
 
Since the pioneering works of Hannes Alfv\'en in the 1930s, ideal magnetohydrodynamics (MHD) has established itself as a cornerstone in fusion and astrophysical plasmas \cite{KT73,GP04,K05}. The ubiquity of ideal MHD stems from its combination of simplicity and (fairly) wide applicability. As MHD is a fluid theory, it shares deep connections with ideal hydrodynamics, including the concept of helicity conservation. Helicities are of considerable interest as they are topological quantities \cite{M69,BF84}, and share a close kinship with the relaxation and self-organization \cite{W58,T74} of plasmas. A third advantage of ideal MHD is that it possesses elegant action principle \cite{Newcomb62} and Hamiltonian \cite{MG80} formulations, each of which has several advantages of its own. 

However, despite its manifold advantages, ideal MHD is \emph{not} a perfect theory, as it fails to take into account two-fluid effects such as electron inertia and the Hall current. To address this issue, a plethora of fluid models have been proposed such as Hall MHD \cite{L60}, electron MHD \cite{RSS91}, inertial MHD \cite{KM14} and extended MHD \cite{S56,L59}. Each of these models originates from the two-fluid model, which is Hamiltonian in nature \cite{SK82}. Yet, several models in the literature have failed to recognize the Hamiltonian nature of extended MHD, thereby giving rise to spurious dissipation; see \cite{KM14} for a discussion of the same. We note that many of the extended MHD models have been derived via an action principle formulation \cite{KLMWW14}, but a Hamiltonian formulation has proven elusive - it was only very recently that a unified Hamiltonian approach to extended MHD was proposed in \cite{HKY14}.

Since several versions of extended MHD exist in the literature, we record here the version we shall analyze,  which,  as shown first by   L\"ust \cite{L59}, can be obtained from an  asymptotically consistent ordering. The extended MHD equations are the following:  the  continuity equation
\begin{equation} \label{ContEq}
\frac{\p \rh}{\p t} + \nabla \cdot \left(\rh {\bf V}\right) = 0,
\end{equation}
the equation for the momentum density
\bq \label{MomDensv1}
\rh \left(\frac{\p {\bf V}}{\p t} + {\bf V}\cdot \nabla {\bf V}\right) = - \nabla p + {\bf J} \times {\bf B}
 - d_e^2  \, {\bf J}\cdot \nabla \left(\frac{{\bf J}}{\rho}\right)\,,  
 \eq
and  Ohm's law
\begin{eqnarray} \label{Ohmv1}
&&{\bf E} + {\bf V} \times {\bf B} - \frac{d_i}{\rho} \, \Big({\bf J} \times {\bf B} - \nabla p_e\Big) \nonumber \\
&&\hspace{3 cm} = \frac{d_e^2}{\rho}\left[\frac{\p {\bf J}}{\p t} + \nabla \cdot \left({\bf V} {\bf J} + {\bf J} {\bf V} 
- \frac{d_i}{\rho}\,  {\bf J} {\bf J} \right) \right].
\end{eqnarray}
Here the variables $\rh$, ${\bf V}$ and ${\bf J}=\nabla \times {\bf B}$ serve as the total mass density, center-of-mass velocity and the current,  respectively,   written in standard Alfv\'en units with $d_e = c/\left(\omega_{pe} L\right)$ and $d_i = c/\left(\omega_{pi} L\right)$ serving  as the electron and ion skin depths normalized by a characteristic  length scale $L$.  As usual,   $c$ is the speed of light and $\omega_{pi,e}$ are the ion and electron plasma frequencies.   In the above equations the total pressure $p$ and the electron pressure $p_e$ will be  assumed to be barotropic, i.e., functions of $\rho$ alone, a consequence of which is that the electron pressure is removed from Ohm's law upon insertion of $\mathbf{E}$ into Faraday's law.   This thermodynamic restriction can be relaxed, but it  will be assumed throughout this paper.    The extended MHD system of Eqs.~\eqref{ContEq}, \eqref{MomDensv1}, and  \eqref{Ohmv1} was shown in \cite{KM14} to conserve the following total energy:
\begin{equation}
 \label{HamExtMHD}
H[d_e;\mathbf{B}] = \int_D d^3x\,\left[\frac{\rh |{\bf V}|^2}{2} + \rh U(\rh) + \frac{|{\bf B}|^2}{2} + d_e^2 \frac{\left|\nabla \times {\bf B}\right|^2}{2\rh} \right]\,, 
\end{equation}
with $U$ being the internal energy function and the pressure given by $p=\rho^2\p U/\p \rho$.  Here the reason for displaying the arguments $[d_e;\mathbf{B}]$ will become clear later.   Observe that the above expression  depends on $d_e$ but is independent of $d_i$. Also  observe that the  last term on the RHS of (\ref{MomDensv1}) that is proportional to $d_e^2$ is  necessary for energy conservation, although it  is often neglected in textbook treatments (see \cite{KM14} for a detailed discussion of this issue).   Two limits of extended MHD will be of interest: the case where $d_e=0$ yields the system commonly referred to as Hall MHD, while we refer to the case where $d_i=0$ as inertial MHD.  

At this stage, it is helpful to recall the many advantages of deriving physical theories via action principles and/or Hamiltonian methods. The former represent an excellent way of building in constraints \emph{a priori} and permit a natural analysis of symmetries and associated invariants via Noether's theorem; we refer the reader to \cite{MLA14,KLMWW14,LM14,LMT14,DMP15} for expositions of this approach in the context of plasmas. The associated Hamiltonian formalism is endowed with several advantages of its own - it enables the determination of a special class of invariants, the Casimirs, which play a crucial role in determining the equilibria and their stability via the energy-Casimir method.  A comprehensive discussion of the Hamiltonian formulation is found in the works of \cite{MG80,pjm82,S88,M98,pjm05,Mor09,AMP10,AMP12,pjmAP13}. 

Our goal in this paper  is twofold in nature. Firstly, we shall demonstrate that Hall, inertial and the full extended MHD models possess  a common underlying structure. We use this commonality to derive Casimir invariants, such as the helicities, through simpler means. After establishing the correspondences between variants of extended MHD in Sections \ref{SecHallInertial} and \ref{SecLagEquiv}, we prove the Jacobi identity for Hall MHD in detail in Section \ref{SecHallJac}. Our work serves as a complement of \cite{HKY14}, where the Hamiltonian structure of extended MHD was analyzed in detail.



\section{On the similarities and equivalences of extended MHD models}
 \label{SecHallInertial}
 
In this section, we analyze Hall MHD and demonstrate its equivalence with inertial MHD and extended MHD. We exploit this equivalence to determine the helicities, which are Casimir invariants, of these models in a straightforward manner.


\subsection{Hall MHD structure} 
\label{SubSecHMHD}

As noted in Sec.~\ref{Intro}, Hall MHD is obtained from extended MHD by setting $d_e=0$.  In Hall MHD, it is assumed that the two species drift with different velocities (as opposed to ideal MHD), but it is assumed that the electrons are inertialess (akin to ideal MHD). We commence our analysis with the Hall MHD bracket  expressed as
\begin{eqnarray} \label{HMHDBrack}
\{F,G\}^{HMHD} &=& - \int_D d^3x\,\Bigg\{\left[F_\rh \nabla \cdot G_{\bf V} + F_{\bf V} \cdot \nabla G_\rh \right] - \left[\frac{\left(\nabla \times {\bf V}\right)}{\rh} \cdot \left(F_{\bf V} \times G_{\bf V}\right)\right] \nonumber \\
&& \quad \, - \left[\frac{{\bf B}}{\rh} \cdot \Big(F_{\bf V} \times \left(\nabla \times G_{\bf B}\right) - G_{\bf V} \times \left(\nabla \times F_{\bf B}\right)\Big)\right] 
\nonumber \\
&&  \quad \, + d_i \left[\frac{{\bf B}}{\rh} \cdot \Big(\left(\nabla \times F_{\bf B}\right) \times \left(\nabla \times G_{\bf B}\right)\Big) \right]\Bigg\},
\end{eqnarray}
where $F_{\bf V}:=\de F/\de {\bf V}$, etc.\  represent the functional derivatives with respect to the corresponding variables. Specifically,  $F_\psi = \de F/\de \psi$ represents the functional derivative with respect to $\psi$, defined via
\begin{equation}
\frac{dF\left[\psi + \epsilon \delta \psi\right]}{d \epsilon}\Bigg|_{\epsilon=0} =: \Bigg<\frac{\de F}{\de \psi},\,\delta \psi\Bigg>\,.
\end{equation}
Before proceeding further, it is worth bearing in mind that the limit $d_i \rightarrow 0$ leads to the ideal MHD bracket first given in \cite{MG80}:
\begin{eqnarray}\label{MHD}
	\{F,G\}^{MHD}&:=&-\int_D d^3x \Bigg(F_{\rho} \mathbf{\nabla} \cdot G_\mathbf{v} -
	G_{\rho} \mathbf{\nabla} \cdot F_\mathbf{v}
	+\frac{\mathbf{\nabla}\times\mathbf{v}}{\rho}\cdot G_\mathbf{v}\times
	F_\mathbf{v}\nonumber\\
	&&   + \frac{\mathbf{B}}{\rho}\cdot\Big\lbrack
	F_\mathbf{v}\cdot\mathbf{\nabla} G_\mathbf{B} -
	G_\mathbf{v}\cdot\mathbf{\nabla} F_\mathbf{B}\Big\rbrack +
	\mathbf{B}\cdot\Big\lbrack \mathbf{\nabla}\frac{F_{\mathbf{v}}}{\rho}\cdot
	G_\mathbf{v}- \mathbf{\nabla}\frac{G_{\mathbf{v}}}{\rho}\cdot
	F_\mathbf{v} \Big\rbrack\Bigg),
\end{eqnarray}

This bracket with the Hamiltonian $H[0;\mathbf{B}]$ of 
\eqref{HamExtMHD} gives the system in the form $\p Z/\p t= \{Z, H\}^{HMHD}$ where $Z=(\rho, \mathbf{V}, \mathbf{B})$ denotes the observables of the model.   For convenience we  re-express (\ref{HMHDBrack}) as 
\begin{equation} 
\label{HMHDOldVar}
\{F,G\}^{HMHD} = \{F,G\}^{MHD} + \{F,G\}^{Hall},
\end{equation}
where $\{F,G\}^{MHD}$ is the ideal MHD bracket given by (\ref{MHD}) and $\{F,G\}^{Hall}$ is the Hall term of (\ref{HMHDBrack}) that involves the ion skin depth $d_i$, first given in  \cite{YH13,HKY14}. We note in passing that an earlier, and alternative, noncanonical bracket for Hall MHD, that uses a redundant variable, was analyzed in \cite{Holm87,KH12}.

Because  $\{F,G\}^{Hall}$ only depends on functional derivatives with respect to $\mathbf{B}$ an immediate consequence is  that \emph{any} Casimir of ideal MHD that is independent of ${\bf B}$ will automatically be  a Casimir of Hall MHD, but it is not necessarily true for ${\bf B}$-dependent Casimirs, such as the magnetic and cross helicities.  However, observe that the magnetic helicity 
\begin{equation} \label{MagHel}
\calc_1 = \int_D d^3x\,{\bf A}\cdot{\bf B}\, ,
\end{equation}
 a Casimir of ideal MHD  also satisfies $\{F,\calc_1\}^{Hall} = 0$. Thus it is indeed true that  (\ref{MagHel}) is also a Casimir of Hall MHD.  A simple calculation shows that the cross helicity $\int_D d^3x\,{\bf v}\cdot{\bf B}$ is not a Casimir of \eqref{HMHDBrack}.

The search for a second helicity led us to introduce a new  magnetic  variable  given by   
\begin{equation} \label{IonCanMom}
\boldsymbol{\calb}_i = {\bf B} + d_i \nabla \times {\bf V}\,.
\end{equation}
Upon effecting the functional chain rule, we obtain our first \underline{remarkable identity}  when  \eqref{HMHDBrack}  is  re-expressed in terms of $\boldsymbol{\calb}_i$:
\begin{equation} \label{HMHDNewVar}
\{\tilde{F},\tilde{G}\}^{HMHD} [d_i;\mathbf{B}] \equiv \{F,G\}^{HMHD}\left[-d_i; {\boldsymbol{\calb}_i}\right] = \{F,G\}^{MHD}\left[{\boldsymbol{\calb}_i}\right] - \{F,G\}^{Hall}\left[{\boldsymbol{\calb}_i}\right],
\end{equation}
where the displayed arguments $d_i$ and  ${\boldsymbol{\calb}_i}$ indicate that the respective components of (\ref{HMHDNewVar}) are the same as (\ref{HMHDOldVar}) except that $d_i$ and ${\bf B}$ are replaced by $-d_i$ and ${\boldsymbol{\calb}_i}$, respectively. The equivalence can be shown by inserting the following chain rule  formulas (see, e.g., \cite{pjm82}):
\bq
 \tilde{F}_{\mathbf{B}}=   F_{\boldsymbol{\calb}_i}
\quad\mathrm{and} \quad 
\tilde{F}_{\mathbf{V}}= {F}_{\mathbf{V}} + d_i\,  \nabla\times {F}_{\boldsymbol{\calb}_i}
\eq
into the left hand side of \eqref{HMHDNewVar}, where we are assuming  that $F[{\boldsymbol{\calb}_i}, \mathbf{V},\rho]= \tilde{F}[\mathbf{B},\mathbf{V},\rho]$. In the above expression(s), observe that the tildes are used to denote the functionals (and the bracket) in terms of the `original' variables, i.e. ${\bf B}$, ${\bf V}$ and $\rh$.

Following the  line of reasoning leading to \eqref{MagHel}, we conclude that
\begin{equation} \label{CanHel}
\calc_2 = \int_D d^3x\,{\boldsymbol{\cala}_i}\cdot{\boldsymbol{\calb}_i} = \left({\bf A} + d_i {\bf V}\right) \cdot \left({\bf B} + d_i \nabla \times {\bf V}\right),
\end{equation}
is a Casimir of Hall MHD.  Again, this follows because it is a Casimir  of ideal MHD, now with ${\bf B}$ replaced by ${\boldsymbol{\calb}_i}$,  and it also satisfies $\{F,\calc_2\}^{Hall}\left[{\boldsymbol{\calb}_i}\right] = 0$.

In summary, the  transformation ${\bf B} \rightarrow {\boldsymbol{\calb}_i}$ exhibits two very special properties: \begin{itemize}
\item We see that it preserves the form of the Hall MHD bracket, i.e. it is evident that (\ref{HMHDOldVar}) and (\ref{HMHDNewVar}) are identical to one another upon carrying out this transformation, apart from the change in sign, viz.  $d_i \rightarrow - d_i$.
\item It allows us to quickly determine the second Casimir of Hall MHD, without going through the conventional procedure of solving a set of constraint equations. In fact, we see that (\ref{MagHel}) and (\ref{CanHel}) possess the same form.  Note that in light of the Casimir $\calc_1$ of \eqref{MagHel},  the second Casimir can be viewed as a sum of  the fluid helicity and cross helicity. 
\end{itemize}
Thus, it is evident that such transformations play a crucial role, both in exposing the symmetries of the system and in determining the Casimirs. In Section \ref{SecLagEquiv}, we shall explore this issue in greater detail. 


\subsection{Hall MHD and inertial MHD} 
\label{SubSecInertHall}

Both ideal MHD and Hall MHD assume that the electrons are inertialess, i.e.\  this is done by taking the limit $m_e/m_i \rightarrow 0$ everywhere. However, there are several regimes where electron inertia effects may be of considerable importance, such as reconnection \cite{OP93}. To address this issue, a new variant of MHD, dubbed inertial MHD, was studied in \cite{KM14} and the Hamiltonian and Action Principle (HAP) formulation of two-dimensional inertial MHD was presented in \cite{LMT14}.

We  now turn our attention to inertial MHD, whose noncanonical bracket is given by
\bqy
\{F,G\}^{IMHD}[d_e; {\bf B}^\star] &=& \{F,G\}^{MHD}\left[{\bf B}^\star\right] 
\nonumber\\
&& \hspace{1 cm} + d_e^2 \int_D d^3x\,\left[\frac{\nabla \times {\bf V}}{\rh}\cdot \Big(\left(\nabla \times F_{{\bf B}^\star}\right) \times \left(\nabla \times G_{{\bf B}^\star}\right) \Big) \right]\, ,
\eqy
where $\{F,G\}^{MHD}\left[{\bf B}^\star\right]$ is  the  ideal MHD bracket (\ref{MHD}) with ${\bf B}$ replaced by ${\bf B}^\star$, where the latter variable given by 
\begin{equation} \label{InertB}
{\bf B}^\star = {\bf B} + d_e^2\,  \nabla \times \left(\frac{\nabla \times {\bf B}}{\rh}\right)\,,
\end{equation}
and represents the `inertial' magnetic field  employed in \cite{LMT14}.   The  variable ${\bf B}^\star $  is approximately the curl of the electron canonical momentum, which has been used in many contexts (e.g.,  \cite{GKR94,CAPBB01}).    The Hamiltonian for this  model is $H[d_e; {\bf B}]$, as given by (\ref{HamExtMHD}), which can also be written as 
\begin{equation}
 \label{HamIMHD}
H[d_e;\mathbf{B}^\star] = \int_D d^3x\,\left[\frac{\rh |{\bf V}|^2}{2} + \rh U(\rh) + \frac{{\bf B}\cdot {\bf B}^\star}{2}  \right]\,,
\end{equation}
where $\mathbf{B}$ is treated as dependent on  ${\bf B}^\star$.

Now consider  the transformation to a new variable
\begin{equation}
\boldsymbol{\calb}_e = {\bf B}^\star - d_e \nabla \times {\bf V},
\end{equation}
and re-express our bracket in terms of it. Upon doing so, we find an interesting result 
\bqy 
\label{IMHDNewVar}
\{\tilde{F},\tilde{G}\}^{IMHD}[d_e;{\bf B}^\star]& \equiv & \{F,G\}^{IMHD}\left[d_e;\boldsymbol{\calb}_e\right]  
\\
&=& \{F,G\}^{MHD}\left[\boldsymbol{\calb}_e\right]  - 2d_e \int_D d^3x\,\left[\frac{\boldsymbol{\calb}_e}{\rh} \cdot \left(\left(\nabla \times F_{\boldsymbol{\calb}_e}\right) \times \left(\nabla \times G_{\boldsymbol{\calb}_e}\right)\right) \right]\,,
\nonumber
\eqy
where, akin to \eqref{HMHDNewVar} and the discussion afterwards, the equivalence is established by the functional chain rule. The second term in the above bracket for inertial MHD can be compared against the last term in (\ref{HMHDBrack}) - we see that the two are identical when $d_i$ is replaced by $2d_e$ and ${\bf B}$ is replaced by $\boldsymbol{\calb}_e$ in the latter expression. Thus, we arrive at our second  \underline{remarkable identity}:
\begin{equation} \label{HallInertEqv}
\{\tilde{F},\tilde{G}\}^{IMHD}[d_e;{\bf B}^\star] \equiv \{F,G\}^{HMHD} \left[2d_e;\boldsymbol{\calb}_e\right].
\end{equation}
In other words, the inertial MHD bracket is equivalent to the Hall MHD bracket when the transformations $d_i \rightarrow 2d_e$ and ${\bf B} \rightarrow \boldsymbol{\calb}_e$ are applied to the latter. As a result, we are led to a series of important  conclusions:
\begin{itemize}
\item As the inertial and Hall MHD brackets are identical under a change of variables (and constants), proving the Jacobi identity for one of them constitutes an automatic proof of the other.
\item The Casimirs of inertial MHD are easily obtained from  the equivalent Casimirs  for Hall MHD, i.e., the following two helicities emerge:
\begin{equation} \label{IMHDHelI}
\calc_I = \int_D d^3x\, \left({\bf A}^\star - d_e {\bf V}\right) \cdot \left({\bf B}^\star - d_e \nabla \times {\bf V}\right),
\end{equation}
\begin{equation} \label{IMHDHelII}
\calc_{II} = \int_D d^3x\, \left({\bf A}^\star + d_e {\bf V}\right) \cdot \left({\bf B}^\star + d_e \nabla \times {\bf V}\right),
\end{equation}
where ${\bf B}^\star = \nabla \times {\bf A}^\star$, and the RHS is determined via (\ref{InertB}). Observe that (\ref{IMHDHelII}) follows from (\ref{CanHel}) upon using the relation (\ref{HallInertEqv}) and the subsequent discussion.
\item By taking the difference of (\ref{IMHDHelII}) and (\ref{IMHDHelI}), we obtain a Casimir:
\begin{equation} \label{IMHDHelIII}
\calc_{III} = \int_D d^3x\, {\bf V}\cdot {\bf B}^\star,
\end{equation}
which is identical to the cross-helicity invariant of ideal MHD, after performing the transformation ${\bf B} \rightarrow {\bf B}^\star$. The existence of this invariant was  documented in \cite{LMT14}.
\end{itemize}
We observe that (\ref{IMHDHelII}) and (\ref{IMHDHelIII}) were previously obtained as the Casimirs for inertial MHD in \cite{HKY14}, and it is thus evident that inertial MHD has not one, but \emph{two} Casimirs (helicities) of the form $\int_D d^3x\,{\bf P} \cdot \left(\nabla \times {\bf P}\right)$, as seen from (\ref{IMHDHelI}) and (\ref{IMHDHelII}). As a result, this allow us to emphasize a rather unique feature of inertial MHD:
\begin{itemize}
\item One can interpret inertial MHD as consisting of two helicities akin to the magnetic (or fluid) helicity, cementing its similarity to Hall MHD and the 2-fluid models \cite{SK82}.
\item Alternatively, we can view inertial MHD as being endowed with one Casimir resembling the magnetic helicity and the other akin to the cross helicity. Such a feature renders it analogous to ideal MHD, which possesses similar features \cite{MG80}.
\end{itemize}
To summarize thus far, we have shown an unusual correspondence between Hall MHD (inertialess, finite Hall drift) and inertial MHD (finite electron inertia, no Hall drift) by showing that the two brackets are equivalent under a suitable set of transformations. We shall explore their origin in more depth in Section \ref{SecLagEquiv}.


\subsection{Extended MHD Casimirs and  interrelations}
 \label{SubSecExtMHD}
 
Hitherto, we have discussed models that separately incorporate the Hall drift and  finite electron inertia. Extended MHD combines these effects together, giving rise to a more complete model. The noncanonical bracket for this model is
\begin{equation} \label{ExMHDBrack}
\{{F},{G}\}^{XMHD} [d_i,d_e; \mathbf{B}^\star]= \{F,G\}^{IMHD}[d_e; \mathbf{B}^\star] + \{F,G\}^{Hall}\left[d_i;\mathbf{B}^\star\right],
\end{equation}
where the second term on the RHS denotes the Hall term with $\mathbf{B}$ replaced by $\mathbf{B}^\star$, and the latter variable was defined in (\ref{InertB}).  This bracket together with  the Hamiltonian  $H[d_e;\mathbf{B}^\star]$ of \eqref{HamIMHD} generates extended MHD. 

It is evident that a clear pattern has emerged:  
\begin{enumerate}
\item The Jacobi identity for the Hall bracket can be proven in a simple manner as it represents the sum of two components, one of which already satisfies the Jacobi identity (the ideal MHD component). The details are provided in Sec.~\ref{SecHallJac}.
\item The Jacobi identity for inertial MHD automatically follows as per the discussion in Section \ref{SubSecInertHall}. 
\item It is easy to see from (\ref{ExMHDBrack}) that the extended MHD bracket will then be composed of a component (inertial MHD) that already satisfies the Jacobi identity, apart from a second component that represents the Hall contribution. As a result, the calculation mirrors the proof of the Jacobi identity for Hall MHD, and the similarities are manifest upon inspecting (\ref{HMHDOldVar}) and (\ref{ExMHDBrack}).  
\end{enumerate}
%
Since we have argued that each of the extended MHD models shares a degree of commonality, it also follows that extended MHD must possess \emph{two} helicities akin to the magnetic helicity (in form), and that they should involve the variables $\mathbf{B}^\star$ and $\mathbf{V}$.  Thus, we seek Casimirs of the form
\begin{equation} \label{HelExtMHD}
\calc_{XMHD} = \int_D d^3x\, \left(\mathbf{V} + \lambda \mathbf{A}^\star\right) \cdot \left(\nabla \times \mathbf{V} + \lambda \mathbf{B}^\star\right),
\end{equation}
and determine $\lambda$ by demanding that $\{{F},\calc_{XMHD}\}^{XMHD}=0$ for all $F$.  This  leads to the following  quadratic equation for $\lambda$: 
\begin{equation} \label{QuadExtMHD}
d_e^2 \lambda^2 + d_i \lambda -1 = 0,
\end{equation}
whose solutions are
\begin{equation}
\lambda = \frac{-d_i \pm \sqrt{d_i^2 + 4d_e^2}}{2d_e^2},
\end{equation}
and these invariants was obtained in \cite{HKY14}. We reiterate the importance of recognizing the existence of \emph{two} helicities akin to the fluid (or magnetic) helicity, since this is a feature that the extended MHD models inherit from the parent two-fluid model. 

In fact, we can recover these two helicities by following the same spirit of variable transformations introduced previously. Hence, we   introduce the variable:
\begin{equation}
\mathbf{B}_\lambda = \mathbf{B}^\star + \lambda^{-1} \nabla \times \mathbf{V},
\end{equation}
where $\lambda$ satisfies (\ref{QuadExtMHD}). Upon doing so, we find our last  \underline{remarkable identity}: 
\begin{equation} \label{ExHallMHDEqv}
\{\tilde{F},\tilde{G}\}^{XMHD}[d_i,d_e; \mathbf{B}^\star]\equiv \{F,G\}^{HMHD}\left[d_i-2\lambda^{-1};\mathbf{B}_\lambda\right], 
\end{equation}
where the RHS indicates that the extended MHD bracket is equivalent to the Hall MHD bracket, when the latter is subjected to the swaps $d_i \rightarrow d_i-2\lambda^{-1}$ and $\mathbf{B} \rightarrow \mathbf{B}_\lambda$. One must bear in mind that there are \emph{two} such variable transformations since there are two choices for $\mathbf{B}_\lambda$ which stem from (\ref{QuadExtMHD}) - the quadratic equation for $\lambda$. We find that these two variable transformations naturally allow us to determine the two helicities of the model. We recover (\ref{HelExtMHD}) successfully, thereby affirming the power of these variable transformations. Furthermore, we conclude from (\ref{ExHallMHDEqv}) that  our proof of the Jacobi identity for Hall MHD  automatically ensures that the extended MHD bracket also satisfies the same property.

In summary, we have established the remarkable result that a proof of the Jacobi identity for the Hall bracket suffices to establish the validity of the inertial and extended MHD brackets as well.  This proof is given in detail in Sec.~\ref{SecHallJac}. 



\section{The Lagrangian origin of the equivalence between the extended MHD models}
 \label{SecLagEquiv}

In this section, we shall briefly explore the origin of the helicities derived in the previous sections, and comment on the equivalences between the various extended MHD models.   In order to do so, we appeal to the Lagrangian picture of fluid models, which envisions the fluid as a continuum collection of particles.  In this picture, laws are built in \emph{a priori} through the imposition of suitable geometric constraints; we refer the reader to \cite{Newcomb62,MLA14,KLMWW14,LMT14} for further details. An extensive treatment will be given in a future publication \cite{eric}.

In ideal MHD, we know that flux is frozen-in, and this translates into a local statement of flux conservation on the Lagrangian level. When one works out the algebra, it is shown that the magnetic induction equation of ideal MHD is just the Lie-dragging of a 2-form - the magnetic field ${\bf B}\cdot d{\bf S}$. Alternatively, one can interpret it, in 3D, as the Lie-dragging of a vector density \cite{LM14,DMP15}. Now, let us take a step back and consider two-fluid theory, where one can define a canonical momentum $\boldsymbol{\calp} = m_s {\bf v}_s + q_s {\bf A}$ for each species. It is evident that ${\bf A}$ represents the electromagnetic component of the canonical momentum, whilst ${\bf v}_s$ gives rise to the kinetic component. Next, suppose that we consider a scenario where the kinetic momentum is much `smaller' than its electromagnetic counterpart - this is achieved especially in the case of electrons, owing to their lower mass. In such an event, we see that the canonical momentum reduces to ${\bf A}$ (up to proportionality factors) and we can interpret ${\bf B}$ as a certain limit of $\nabla \times \boldsymbol{\calp}$. In ideal MHD, which is a pure one-fluid theory, it is easy to view ${\bf B}$ as being Lie-dragged by the center-of-mass velocity ${\bf V}$. 

Now, we shall proceed in the same heuristic manner, through the incorporation of two-fluid effects. Firstly, let us suppose that the electrons are inertialess, but \emph{not} the ions. As a consequence, one finds that the center-of-mass velocity ${\bf V}$ and the ion velocity virtually coincide. The corresponding canonical momenta, after suitable normalization, reduce to ${\bf B}$ and $\boldsymbol{\calb}_i$ respectively, after rewriting them in terms of one-fluid variables. Following the analogy outlined above, we can choose to Lie-drag them as 2-forms, akin to the magnetic field in ideal MHD. Next, the question arises: by which velocity must we Lie-drag these variables?    The answer is intuitive: we choose to Lie-drag them by the velocity of the corresponding species.  After some manipulation, it is easy to show that the resulting equations are equivalent to those of Hall MHD. 

Next, suppose that we include the effects of electron inertia. The curls of the canonical momenta, when written in terms of the one-fluid variables, are closely connected to ${\bf B}^\star \pm d_e \nabla \times {\bf V}$, and the latter are the variables that appear in (\ref{IMHDHelII}) and (\ref{IMHDHelI}) respectively. Following the same prescription, we can choose to Lie-drag these quantities. 
We choose to Lie-drag the two variables presented above, related to the canonical vorticities, by suitable flow velocities, ${\bf V} \pm d_e \nabla \times {\bf B}/\rh$, which are determined by an appropriate manipulation of the inertial MHD equations. It is found, after some algebraic simplification, that the resulting equations are equivalent to those of inertial MHD. The generalization to extended MHD is not entirely straightforward, but it can be done by using the variables from (\ref{HelExtMHD}) as Lie-dragged 2-forms, and noting that each is Lie-dragged by the effective velocity of the corresponding species.

Thus, we see that our preceding analysis establishes two very important points. Firstly, the equations for extended MHD can be viewed as the natural manifestation of underlying (Lagrangian) geometric constraints.  Secondly, we see that the variables $\boldsymbol{\calb}_i$, $\boldsymbol{\calb}_e$, etc. introduced earlier, and the helicities of the models, are also `natural' --  they emerge from the  unifying concept that all extended MHD models possess two Lie-dragged 2-forms (which share close connections with the canonical momenta). In both these aspects, we see that the Lagrangian picture of extended MHD presents a compelling argument as to why the variable transformations of Section \ref{SecHallInertial} are \emph{not} arbitrary, and, more importantly, it emphasizes the underlying geometric nature of the extended MHD models. The latter is all the more useful as it further serves to emphasize the existence of a unifying structure for the extended MHD models.  The details of this intuitive picture are not trivial and, as noted above,  will be given  in a future publication \cite{eric}.



\section{Jacobi identity for Hall MHD} 
\label{SecHallJac}

In this section, we  present a detailed proof of the Jacobi identity for the noncanonical Hall MHD bracket. The transformations discussed in the preceding sections ensures the Jacobi identity for other versions of extended MHD.

In the absence of the Hall term, we see that (\ref{HMHDBrack}) reduces to the ideal MHD bracket, first obtained in \cite{MG80}, and given by (\ref{MHD}), which is known to satisfy Jacobi identity on its own \cite{MG80,pjm82}.  To reduce clutter we will use the convention throughout that the $\mathbf{\nabla}$ operator acts only on the
variable immediately following it and dyadics will be written as  follows:
\begin{equation}
	\mathbf{B}\cdot\mathbf{\nabla}\frac{F_\mathbf{v}}{\rho}\cdot G_\mathbf{v} =
	B_i\,\partial_i\left(\frac{F_v^j}{\rho}\right)G_v^j.
\end{equation}


\subsection{Hall - Hall Jacobi identity}
 \label{SubSecHallHallJac}
 
Introduction of the Hall current lead  to the additional Hall term   given in  (\ref{HMHDOldVar}), 
\begin{eqnarray}\label{Hall}
	\{F,G\}^{Hall}:= - d_i\int_D d^3 x \frac{\mathbf{B}}{\rho}\cdot\Big\lbrack
	\left(\mathbf{\nabla}\times F_\mathbf{B}\right)\times\left(\mathbf{\nabla}\times G_\mathbf{B}\right)
	\Big\rbrack,
\end{eqnarray}
Demonstrating that Hall MHD bracket satisfies Jacobi is important since it is closely connected to the rest of the extended MHD models, as discussed previously. The
Jacobi identity involves proving that cyclical permutations of any functionals $F,G,H$ vanish, i.e. we require
\begin{equation}\label{Jacobi}
	0 =
	\{\{F,G\},H\}+\{\{G,H\},F\}+\{\{H,F\},G\}\equiv\{\{F,G\},H\}
	+ \substack{\circlearrowleft \\ {F,G,H}}
\end{equation}
Here $\{,\} := \{,\}^{MHD} + \{,\}^{Hall}$ . Because we already know that~\eqref{MHD} satisfies Jacobi and according to the bilinearity of Poisson brackets, the general proof splits into two pieces
\begin{equation}\label{JacobiHMHD}
	\{\{F,G\}^{MHD},H\}^{Hall} + \{\{F,G\}^{Hall},H\}^{MHD}
	+ \substack{\circlearrowleft \\ {F,G,H}} = 0,
\end{equation}
and
\begin{equation}\label{JacobiH}
	\{\{F,G\}^{Hall},H\}^{Hall}
	+ \substack{\circlearrowleft \\ {F,G,H}} = 0.
\end{equation}
This split occurs since (\ref{JacobiHMHD}) involves terms that are linear in $d_i$, whilst (\ref{JacobiH}) is quadratic in $d_i$. We introduce the cosymplectic operator $J$ which depends on the field variables $u$ in general. It is known that Poisson brackets can be formally written in the form
\begin{equation}
	\{F,G\} := \Big<\frac{\delta F}{\delta u}\Big| J \frac{\delta G}{\delta u}\Big>.
\end{equation}
The outer brackets in both~\eqref{JacobiHMHD} and~\eqref{JacobiH} require evaluation of the variational derivatives of the inner bracket with respect to the field variables:
\begin{eqnarray}
	\frac{d}{d\epsilon}\{F,G\}\lbrack u + \epsilon \delta u \rbrack
	\Big|_{\epsilon = 0} &:=& \Big< \frac{\delta}{\delta u}\{F,G\}\Big|\delta u\Big>
	= \Big<\frac{\delta^2 F}{\delta u\delta u} \delta u \Big| J \frac{\delta
	G}{\delta u} \Big>
	\nonumber \\
 && \hspace{ 1 cm} + \Big<\frac{\delta F}{\delta u} \Big|
 J \frac{\delta^2 G}{\delta u\delta u} \delta u \Big> + \Big<\frac{\delta
 F}{\delta u}\Big|\frac{\delta J}{\delta u} (\delta u) \frac{\delta G}{\delta
 u}\Big>
\end{eqnarray}
Proving the Jacobi identity for noncanonical Poisson brackets is aided by a theorem proven in \cite{pjm82}, which states that the contributions from first two terms of the above expression vanish upon cyclic permutation when plugged in the outer bracket.   Thus, we can neglect second variations that appear throughout the following calculations.  Since the outer Hall bracket only  involves variations with respect to  $\mathbf{B}$, it is enough to consider
\begin{equation}\label{HallB}
	\frac{\delta}{\delta\mathbf{B}}\{F,G\}^{Hall}=-d_i \left(\mathbf{\nabla}\times
	F_\mathbf{B}\right)\times\left(\mathbf{\nabla}\times G_\mathbf{B}\right) = -d_i\,
	F_\mathbf{A}\times G_\mathbf{A}\, 
\end{equation}
where the  equalities above are  modulo the second variations  that have been dropped and in the second equality we have introduced the shorthand
\bq
F_\mathbf{A}=\mathbf{\nabla}\times F_\mathbf{B} \,.
\label{FA}
\eq
From \eqref{FA} it follows that  $\nabla \cdot F_{\bf A} = 0$. Substituting~\eqref{HallB} into the Hall-Hall part of the Jacobi relation~\eqref{JacobiH} gives
\begin{equation}
	d_i^2\int_D d^3 x\,
    \mathbf{B}\cdot\Bigg(\mathbf{\nabla}\Big(\frac{1}{2\rho^2}\Big)\times\Big\lbrack
   F_\mathbf{A}\times G_\mathbf{A}	\Big\rbrack
   +\frac{1}{\rho^2}\mathbf{\nabla}\times\Big\lbrack F_\mathbf{A}\times
   G_\mathbf{A} \Big\rbrack\Bigg)\times H_\mathbf{A}.
\end{equation}
Using the vector identities such as
$\mathbf{X}\times\left(\mathbf{Y} \times \mathbf{Z}\right) = \mathbf{Y}\,
\left(\mathbf{X}\cdot\mathbf{Z}\right) - \mathbf{Z}\,\left(\mathbf{X}\cdot \mathbf{Y}\right)$ and $\mathbf{\nabla}\times\left(\mathbf{X}\times\mathbf{Y}\right)
= \mathbf{\nabla}\cdot\left(\mathbf{Y}\,\mathbf{X} -
\mathbf{X}\,\mathbf{Y}\right)$,  collecting similar terms together, permuting,   and integrating by parts, we arrive at
\begin{eqnarray}
	\{\{F,G\}^{Hall},H\}^{Hall}
	+ \substack{\circlearrowleft \\ {F,G,H}} &=& d^2_i\int_D d^3 x\,
	\rho^{-2}
	(F_\mathbf{A}\times G_\mathbf{A}) \cdot
	(H_\mathbf{A}\cdot\mathbf{\nabla}) \mathbf{B}+ \substack{\circlearrowleft \\
	{F,G,H}} \nonumber \\
	&=& d_i^2\int_D d^3 x\, \rho^{-2}\, \epsilon_{ijk} F_A^j\,G_A^k H_A^l \partial_l
	B^i+ \substack{\circlearrowleft \\
	{F,G,H}} \nonumber \\
	&=& d_i^2\int_D d^3 x\, \rho^{-2}\, {F_\mathbf{A}\cdot (
	G_\mathbf{A}\times H_\mathbf{A})}\,  \delta^l_i  \partial_l B^i + \substack{\circlearrowleft \\
	{F,G,H}},
\end{eqnarray}
where the last step becomes apparent when we explicitly write down the other two permutations and use the antisymmetry of Levi-Civita tensor $\epsilon_{ijk}$ in addition to the identity $\epsilon_{ijk}\epsilon^{ljk} = 2\delta^l_i$. Finally, upon invoking the identity $\mathbf{\nabla \cdot
B} = 0$, we see that the Hall term of the bracket \eqref{HMHDBrack} alone satisfies the Jacobi  identity.


\subsection{Hall -  MHD Jacobi identity} 
\label{SubSecHallMHDJac}

We observe that this part is harder to tackle, owing to the greater complexity of the resultant expression.   First consider the first term of \eqref{JacobiHMHD}.  As described in the previous section, the outer Hall bracket~\eqref{Hall} necessitates only the explicit variational derivatives with
respect to $\mathbf{B}$. Hence, we only need to consider such variations of the inner MHD bracket~\eqref{MHD}:
\begin{equation}
	\frac{\delta}{\delta \mathbf{B}}\{F,G\}^{MHD} =
	-\frac{F_\mathbf{v}}{\rho}\cdot\mathbf{\nabla}\,G_\mathbf{B}+
	\frac{G_\mathbf{v}}{\rho}
	\cdot\mathbf{\nabla}\,F_\mathbf{B}-
	\mathbf{\nabla}\frac{F_\mathbf{v}}{\rho}\cdot G_\mathbf{B}+
	\mathbf{\nabla}\frac{G_\mathbf{v}}{\rho}\cdot F_\mathbf{B}+\dots,
\end{equation}
and we have suppressed the implicit second-order variations, as they do not contribute to the Jacobi identity. After substitution into the outer Hall bracket, we get
\begin{eqnarray}
	\{\{F,G\}^{MHD},H\}^{Hall} + \substack{\circlearrowleft \\ {F,G,H}} &=&
	-d_i\int_D \frac{\mathbf{B}}{\rho}\cdot\Bigg\lbrack
	\mathbf{\nabla}\times\bigg(
	\frac{F_\mathbf{v}}{\rho}\cdot\mathbf{\nabla}\,G_\mathbf{B}-
	\frac{G_\mathbf{v}}{\rho}
	\cdot\mathbf{\nabla}\,F_\mathbf{B}\nonumber\\
	&&  \hspace{-.5 cm}+ \mathbf{\nabla}\frac{F_\mathbf{v}}{\rho}\cdot G_\mathbf{B}-
	\mathbf{\nabla}\frac{G_\mathbf{v}}{\rho}\cdot
	F_\mathbf{B}\bigg)\times(\mathbf{\nabla}\times H_\mathbf{B})\Bigg\rbrack +
	\substack{\circlearrowleft \\ {F,G,H}} \,,
\end{eqnarray}
which upon using the vector  identities $\mathbf{\nabla}\times\mathbf{\nabla} f = 0$ and
$\mathbf{X}\times\mathbf{\nabla} \times \mathbf{Y} =
\mathbf{\nabla} \mathbf{Y}\cdot\mathbf{X} - \mathbf{X}\cdot\mathbf{\nabla}
\mathbf{Y}$  simplifies to the following  expression: 
\begin{equation} \label{T2P1}
	\{\{F,G\}^{MHD},H\}^{Hall}  = -d_i\int_D d^3x\,
	\frac{\mathbf{B}}{\rho}\cdot\left(\mathbf{\nabla}\times\frac{F_\mathbf{v}\times
	G_\mathbf{A}-G_\mathbf{v}\times F_\mathbf{A}}{\rho}\times H_\mathbf{A} \right). 
\end{equation}

Now consider the  second term of~\eqref{JacobiHMHD}.  The outer MHD bracket of this term requires evaluation of variations with respect to both $\mathbf{B}$ and $\rho$. We
already have the first one from~\eqref{HallB}, while the second yields
\begin{equation}
	\frac{\delta}{\delta \rho}\{F,G\}^{Hall} =  d_i\frac{\mathbf{B}}{\rho^2} \cdot
	(F_\mathbf{A}\times G_\mathbf{A})\,.
	\label{Hrho}
\end{equation}
Upon substituting \eqref{HallB} and \eqref{Hrho} into the second term of~\eqref{JacobiHMHD}, we end up with
\begin{equation} \label{T2P2}
	-d_i\int_D d^3x\, \frac{\mathbf{B}}{\rho^2}\cdot \left(F_\mathbf{A}\times G_\mathbf{A}\right)
	 \mathbf{\nabla} \cdot H_\mathbf{v} +
	\frac{\mathbf{B}}{\rho}\cdot\left[\left(\mathbf{\nabla}\times\frac{F_\mathbf{A}\times
	G_\mathbf{A}}{\rho}\right)\times H_\mathbf{v}\right] +
	\substack{\circlearrowleft \\ {F,G,H}} 
\end{equation}
Upon combining (\ref{T2P1}) and (\ref{T2P2}), we have
\begin{eqnarray}\label{JacobiPre}
	\calj &=& -d_i\int d^3 x\Bigg( \frac{\mathbf{B}}{\rho^2}\cdot (F_\mathbf{A}\times
	G_\mathbf{A}) \mathbf{\nabla} \cdot H_\mathbf{v} +
	\frac{\mathbf{B}}{\rho}\cdot\left[\left(
	\mathbf{\nabla}\times\frac{F_\mathbf{A}\times
	G_\mathbf{A}}{\rho}\right)\times H_\mathbf{v}\right] \nonumber
	\\
&& \quad \quad	+ \frac{\mathbf{B}}{\rho}\cdot\left[\left(\mathbf{\nabla}\times\frac{F_\mathbf{v}\times
	G_\mathbf{A}-G_\mathbf{v}\times F_\mathbf{A}}{\rho}\right)\times H_\mathbf{A}
	\right]\Bigg) +
	\substack{\circlearrowleft \\ {F,G,H}} \nonumber \\
	&=& \calj_1 + \calj_2 + \calj_3,
\end{eqnarray}
where $\calj_i$'s represent the three contributions arising from  (\ref{T2P2}) and (\ref{T2P1}),  respectively. Applying the vector identities mentioned previously, and recollecting that variations with respect to $\mathbf{A}$ are divergence-free, the third term can be manipulated to yield
\begin{eqnarray}
	\calj_3 &=& d_i\int_D d^3 x\, \frac{\mathbf{B}}{\rho} \cdot
	H_\mathbf{A}\times\left(G_\mathbf{A}\cdot\mathbf{\nabla}\frac{F_\mathbf{v}}{\rho}
	- \mathbf{\nabla}\cdot
	F_\mathbf{v}\frac{G_\mathbf{A}}{\rho}-F_\mathbf{v}\cdot\mathbf{\nabla}
	\frac{G_\mathbf{A}}{\rho}\right.
	\nonumber\\
	&&\hspace{6.25cm} \left. -  
	F_\mathbf{A}\cdot\mathbf{\nabla}\frac{G_\mathbf{v}}{\rho}
	+ \mathbf{\nabla}\cdot
	G_\mathbf{v}\frac{F_\mathbf{A}}{\rho}+G_\mathbf{v}\cdot\mathbf{\nabla}
	\frac{F_\mathbf{A}}{\rho}\right)\nonumber \\ 
	&=& -d_i\int_D d^3 x \left( -\frac{2\mathbf{B}}{\rho^2}\cdot \left(F_\mathbf{A}
	\times G_\mathbf{A}\right)  \mathbf{\nabla}\cdot H_\mathbf{v} 
	- \mathbf{B}\cdot \left(F_\mathbf{A}\times G_\mathbf{A}\right) 
	 H_\mathbf{v}\cdot \mathbf{\nabla}
	\left(\frac{1}{\rho^2}\right) \right. \label{j3}
	\\ 
	&&  \hspace{.3cm} -
	\left. \mathbf{B}\cdot\Big(\frac{H_\mathbf{v}}{\rho^2}\cdot\mathbf{\nabla} \Big)
	\Big(F_\mathbf{A}\times G_\mathbf{A}\Big) 
	+ 
	\frac{\mathbf{B}}{\rho}\cdot\Big\lbrack
	F_\mathbf{A}\times(G_\mathbf{A}\cdot\mathbf{\nabla}) -
	G_\mathbf{A}\times(F_\mathbf{A}\cdot\mathbf{\nabla})\Big\rbrack
	\frac{H_\mathbf{v}}{\rho}\right)\,.
	\nonumber
\end{eqnarray}
Here, as before, we have  permuted $F,G,H$. When \eqref{j3} is combined with   $\calj_1$, this results in
\bqy
\label{J1J3sum}
	\calj_1 + \calj_3 &=& d_i\int_D d^3 x \,\bigg(
	\mathbf{\nabla}\cdot\left[
	\frac{H_\mathbf{v}}{\rho^2} (F_\mathbf{A}\times  G_\mathbf{A})\right] \cdot\mathbf{B} 
	\nonumber\\
	 && \hspace {1cm} -
	\frac{\mathbf{B}}{\rho}\cdot\Big\lbrack
	F_\mathbf{A}\times(G_\mathbf{A}\cdot\mathbf{\nabla}) -
	 G_\mathbf{A}\times(F_\mathbf{A}\cdot\mathbf{\nabla})\Big\rbrack
	\frac{H_\mathbf{v}}{\rho} \bigg).
\eqy
The second term of~\eqref{JacobiPre} can be rewritten as
\begin{equation} \label{J2mod}
	\calj_2=-d_i \int_D d^3 x \Bigg(\frac{H_\mathbf{v}}{\rho}\cdot\mathbf{\nabla}
	\left(\frac{F_\mathbf{A}\times G_\mathbf{A}}{\rho}\right) \cdot\mathbf{B} -
	\mathbf{B}\cdot\mathbf{\nabla} \left(\frac{F_\mathbf{A}\times G_\mathbf{A}}{\rho}\right)
	\cdot \frac{H_\mathbf{v}}{\rho} \Bigg)\,.
\end{equation}
Upon using (\ref{J1J3sum}) and (\ref{J2mod}), we can condense~\eqref{JacobiPre} into
\begin{eqnarray}
	\calj &=& d_i \int_D d^3 x \Bigg(\mathbf{B}\cdot\left(\frac{F_\mathbf{A}\times
	G_\mathbf{A}}{\rho}\right)   \mathbf{\nabla}\cdot\left(\frac{H_\mathbf{v}}{\rho}\right)  -
	\mathbf{B}\cdot\mathbf{\nabla} \left(\frac{H_\mathbf{v}}{\rho}\right)
	\cdot\left( \frac{F_\mathbf{A}\times G_\mathbf{A}}{\rho} \right)
	\nonumber \\
&& \hspace{3 cm}  -
	\frac{\mathbf{B}}{\rho}\cdot\bigg[
	F_\mathbf{A}\times(G_\mathbf{A}\cdot\mathbf{\nabla}) -
	G_\mathbf{A}\times(F_\mathbf{A}\cdot\mathbf{\nabla})\bigg]
	\frac{H_\mathbf{v}}{\rho} \Bigg).
	\label{J}
\end{eqnarray}
The second term has been integrated by parts and use has been made of 
$\mathbf{\nabla}\cdot \mathbf{B} = 0$ to obtain \eqref{J}. Without further permutations of $F$, $G$ and $H$, it  can be shown that the first two and the last two terms collapse into
\begin{equation}
	\calj = d_i \int_D d^3 x
	\Bigg(\frac{\mathbf{B}}{\rho}\cdot\left[\Big((F_\mathbf{A}\times
	G_\mathbf{A})\times
	\mathbf{\nabla}\Big)\times\frac{H_\mathbf{v}}{\rho}\right] -
\frac{\mathbf{B}}{\rho}\cdot\left[\Big((F_\mathbf{A}\times
	G_\mathbf{A})\times
	\mathbf{\nabla}\Big)\times\frac{H_\mathbf{v}}{\rho}\right]\Bigg) \equiv 0\,.
\end{equation}
This follows immediately by the application of Lemma 2 from the Appendix of \cite{M13}.  As a result, we see that the Hall - MHD Jacobi identity is satisfied. 

Hence, from the results derived in Sections \ref{SubSecHallHallJac} and \ref{SubSecHallMHDJac}, we conclude that the Hall MHD bracket (\ref{HMHDBrack}) satisfies the Jacobi identity, thereby rendering it a valid noncanonical Poisson bracket. In turn, this ensures the validity of the inertial MHD bracket and, by invoking the identity \eqref{ExHallMHDEqv}, it follows that the extended MHD bracket also satisfies the Jacobi identity.



\section{Conclusion} 
\label{SecConc}

The construction of valid noncanonical Poisson brackets for arbitrary field theories can be challenging, owing to two reasons. Most brackets are constructed through guesswork, and this is a difficult task when confronted with a complex model. Secondly, the task of proving the Jacobi identity is an onerous one, involving a high degree of tedious algebra. 

In this paper, we have bypassed some of these difficulties for extended MHD noncanonical Poisson brackets by primarily appealing to the commonality in the structure of the different extended MHD models. In Section \ref{SecHallInertial}, we showed that certain variable transformations left the form of Hall MHD unchanged, and enabled us to extract the Casimirs (helicities) of the model. In the same spirit, we proved the existence of remarkable changes of variables that highlighted the existence of a common bracket for inertial,  Hall, and extended  MHD. We also commented on the high degree of overlap between these models, and extended MHD in its entirety. 

In the subsequent section, we briefly appealed to  the Lagrangian formulation of fluid models to intuitively  trace the origin of these variable transformations and the emergence of two fluid-like helicities for all extended MHD models, leaving the full analysis for a future publication.   We argued  that each of these models could be  endowed with suitable Lie-dragged 2-forms (and their corresponding flow velocities), which gave  rise to the correct dynamical equations, and ensured the conservation of helicities. Thus, we demonstrated that the commonality of the extended MHD models is a natural consequence of the underlying geometric constraints. We believe that this constitutes an excellent example of the synergy between geometry and physics, and one that will  be exploited for further use in subsequent works.

Lastly, we returned to Hall MHD and presented a detailed proof to show that its bracket, exemplified by (\ref{HMHDBrack}), satisfies the Jacobi identity. As we established that the extended MHD models are closely connected to each other, we believe that this proof will help clarify and corroborate the results derived in \cite{HKY14}.

\begin{acknowledgments}
We would like to acknowledge useful comments by H.\  Abdelhamid, Y.\  Kawazura,  Z.\  Yoshida, S.\ J.\ Benavides and F.\  Pegoraro. 
This research was supported by U.S. Dept.\ of Energy Contract \# DE-FG02-04ER54742.

\end{acknowledgments}


%

\end{document}